\documentclass{iauc}
\usepackage{graphicx}  
\usepackage{natbib}
\pubyear{2005}
\volume{199}
\pagerange{1--}
\jname{Probing Galaxies through Quasar Absorption Lines}
\editors{P. R. Williams, C. Shu, and B. M\'{e}nard, eds.}
\def\simless{\mathbin{\lower 3pt\hbox
   {$\rlap{\raise 5pt\hbox{$\char'074$}}\mathchar''7218$}}}   
\def\simgreat{\mathbin{\lower 3pt\hbox
   {$\rlap{\raise 5pt\hbox{$\char'076$}}\mathchar''7218$}}}   
\def\apj{{ApJ}}

\def\mnras{MNRAS}

\title[Ionization front interactions with density fluctuations]
{Ionization fronts and their interaction with density fluctuations: implications for 
reionization} 
\author[I.~T.~Iliev et al.]
{Ilian T. Iliev$^1$, Paul R. Shapiro$^2$, Evan Scannapieco$^3$, 
Garrelt Mellema$^4$,  Marcelo Alvarez$^2$ 
Alejandro C. Raga$^5$, and Ue-Li Pen$^1$}
\affiliation{$^1$ Canadian Institute for Theoretical Astrophysics, 
University of Toronto, 60 St. George Street, Toronto, ON M5S 3H8, Canada, email:
iliev@cita.utoronto.ca\\[\affilskip]
$^2$ Department of Astronomy, University of Texas, Austin, 78712, USA\\[\affilskip]
$^3$  Kavli Institute for Theoretical Physics,
     Kohn Hall, UC Santa Barbara, Santa Barbara, CA 93106, USA\\[\affilskip]
$^4$ ASTRON, P.O. Box 1, NL-7990 AA Dwingeloo, The
Netherlands\\[\affilskip]
$^5$ Instituto de Ciencias Nucleares,
       Universidad Nacional Autonoma de M\'exico (UNAM),
       Apdo. Postal 70-543, 04510 M\'exico, D. F., M\'exico
}

\begin{document}
\maketitle              

\begin{abstract}

The propagation of cosmological ionization fronts (I-fronts) during reionization 
is strongly influenced by small-scale structure. Here we summarize our recent 
attempts to understand the effect of this small-scale structure. We present high 
resolution cosmological N-body simulations at high-z ($z>6$) which resolve a wide 
range of halo mass, from minihalos to clusters of large, rare halos. We also 
study how minihalos affect I-fronts, through simulations of minihalo photoevaporation
by numerical gasdynamics with radiative transfer. 
Furthermore, we modify the I-front propagation equations to account for evolving 
small-scale structure, and incorporate these results into a semi-analytical 
reionization model. When intergalactic medium clumping and minihalo clustering 
around sources are included, small-scale structure affects reionization by slowing 
it down and extending it in time. This helps to explain observations by the Wilkinson 
Microwave Anisotropy Probe, which imply an early and extended reionization epoch. 
We also study how source clustering affects the evolution and size of H~II regions, 
finding, in agreement with simulations, that H~II regions usually expand, rarely 
shrinking. Hence, ``relic H II regions'' are an exception, rather than the rule. 
When the suppression of small-mass sources in already-ionized regions by Jeans-mass 
filtering is accounted for, H~II regions are smaller, delaying overlap.  
We also present a new numerical method for radiative transfer which is fast, 
efficient, and easily coupled to hydrodynamics and N-body codes, along with sample 
tests and applications. 

\end{abstract}

\section{Introduction}
When a source of ionizing radiation turns on in a neutral gas, the I-front 
initially moves supersonically as a weak, R-type front and outruns the gas 
dynamical disturbance it creates  \citep[e.g.][and refs. therein]{1978ppim.book.....S}.
In a uniform gas, the front eventually slows to twice the isothermal sound 
speed of the ionized gas as it approaches the size of the equilibrium Str\"omgren 
sphere and is transformed into a subsonic D-type front, usually preceded by a shock.
Thereafter, it is very much affected by gas dynamics. \citet{1987ApJ...321L.107S} 
first considered this problem in the cosmological context of the expanding IGM, 
deriving the general equations describing I-front evolution and solving them 
analytically for the case of constant gas clumping factor and temperature. They 
found that global cosmological I-fronts typically stayed in the weak, R-type regime
as the density dropped over time, failing to reach the Str\"omgren radius. This 
result justifies subsequent approximate treatments of cosmic reionization in the 
inhomogeneous IGM that results from structure formation which neglect the gasdynamical
back reaction of the I-fronts. However, when the I-fronts encounter self-shielding
inhomogeneities (like halos) large and dense enough to trap the I-fronts and convert 
them to D-type, this approximation must break down.

\section{Structure formation at high redshift and its effects on I-front
  propagation during reionization}

Within the CDM paradigm cosmological structures form hierarchically in time,
starting with the smallest-scale structures at the highest redshifts, which 
grow through accretion and mergers to form larger and larger structures. While
the cold dark matter clumps gravitationally on all scales, only virialized 
halos with total mass above the cosmological Jeans mass can retain their gas
content. These virialized halos can be divided into two main classes, halos with
virial temperature above $\sim10^4$ K whose gas can cool through atomic line
cooling of hydrogen, leading to efficient star formation in such halos, and
smaller-mass halos (minihalos) whose gas in absence of metals can only cool 
through $H_2$ molecules, which are fragile and easily destroyed. Thus, the 
first class of halos are generally believed to be the dominant sources of the 
radiation responsible for cosmic reionization, while typical minihalos were 
unable to form stars once the first generation created a background of 
dissociating UV starlight. Subsequently, minihalos were just inert balls of 
neutral gas. 
\begin{figure}
 \includegraphics[width=2.7in]{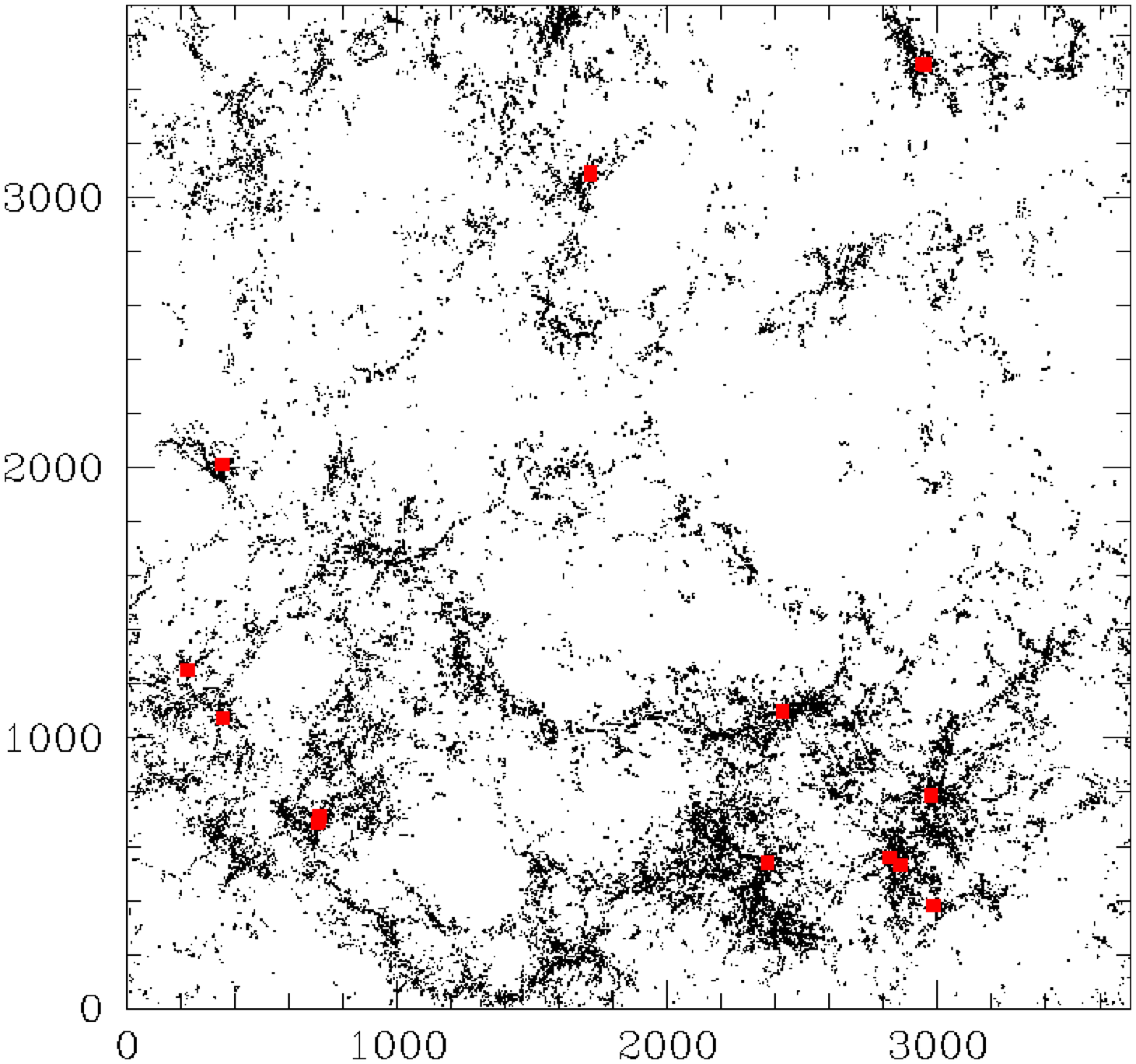}
 \includegraphics[width=2.7in]{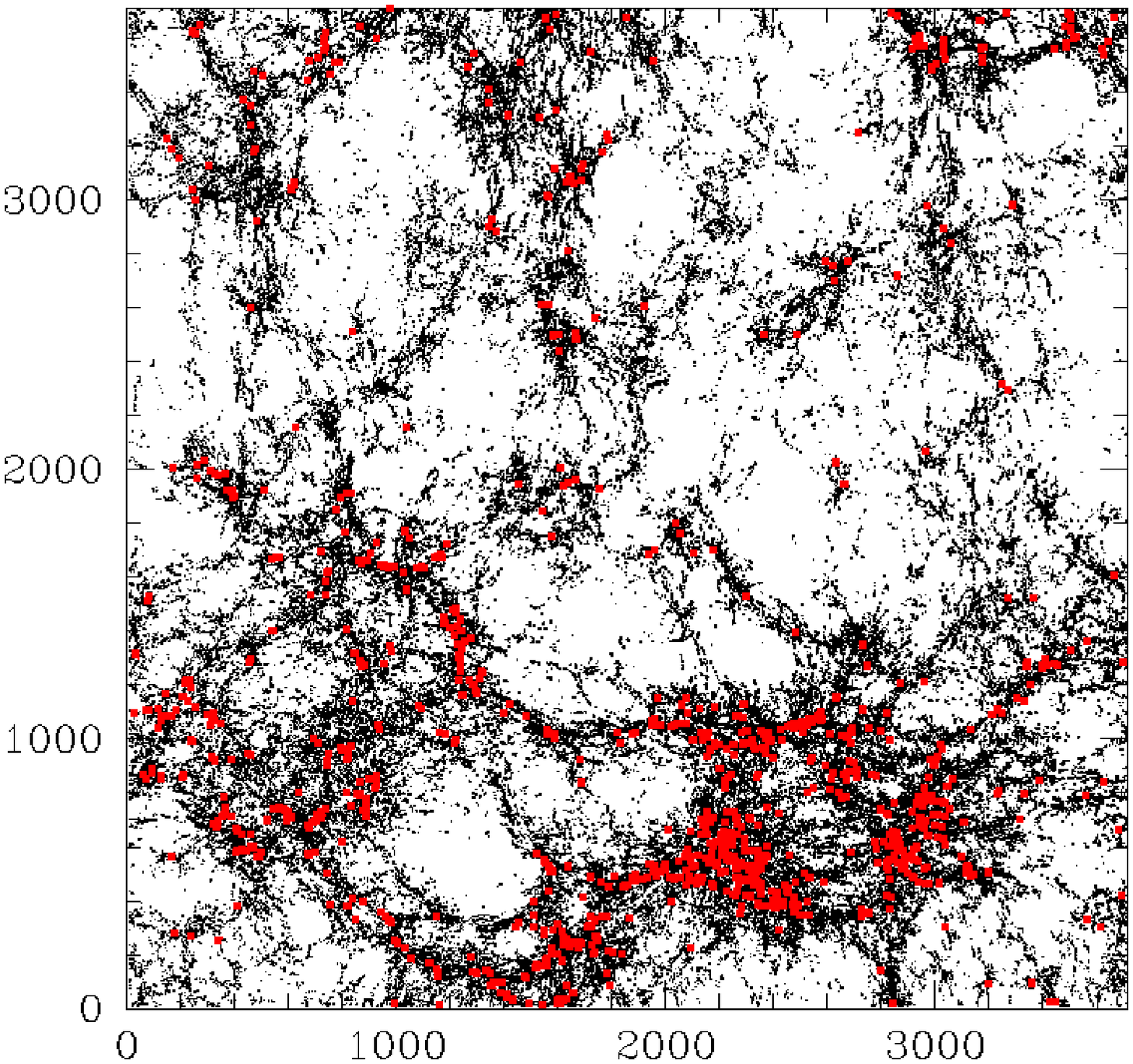}
  \caption{Structure formation at high redshift in $\Lambda$CDM. Halo distribution
at (left) redshift $z=17.2$ and (right) $z=9.42$ from massively-parallel 
PMFAST N-body simulation with $1836^3$ particles, $3672^3$ cells, simulation box 
size $10h^{-1}$ Mpc comoving. Shown are
minihalos (black dots) and ionizing sources (red squares) in a 1 Mpc thick
slice (approx. 1/14th of the total simulation volume).
\label{pmfast_figs}
\vspace{-0.4cm}
}
\end{figure}
Figure~\ref{pmfast_figs} shows the halos which formed in a large N-body 
simulation  of early structure formation. The 
larger (source) halos (red squares) are quite rare and highly
clustered, especially at high redshift, as expected for rare density peaks 
in a Gaussian density field. On the other hand, the minihalos (black dots) 
are quite common at all times and are strongly clustered around
the ionizing sources, covering their sky \citep{2001AIPC..586..219S}. 

\begin{figure}
\includegraphics[width=2.7in]{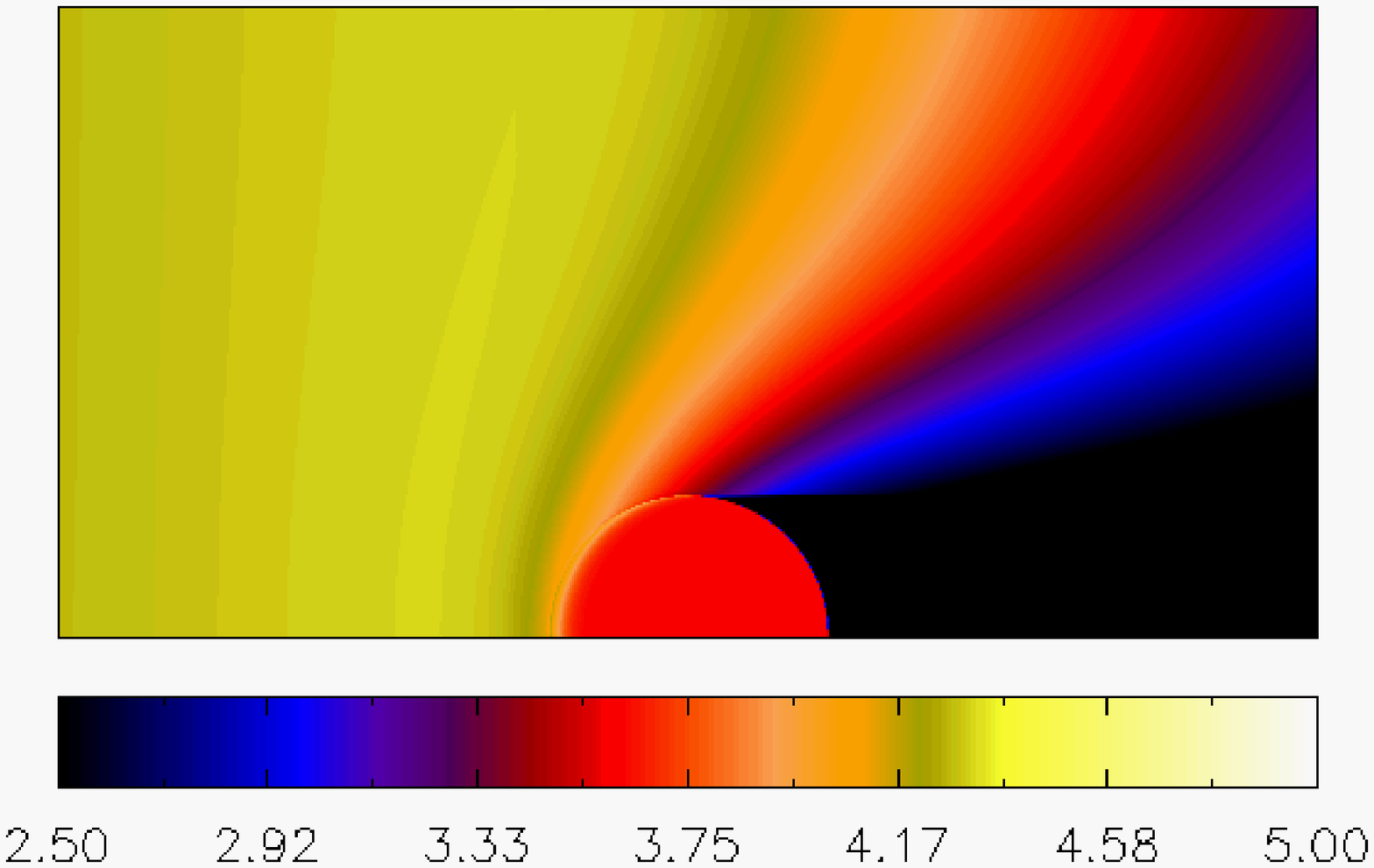}
\includegraphics[width=2.7in]{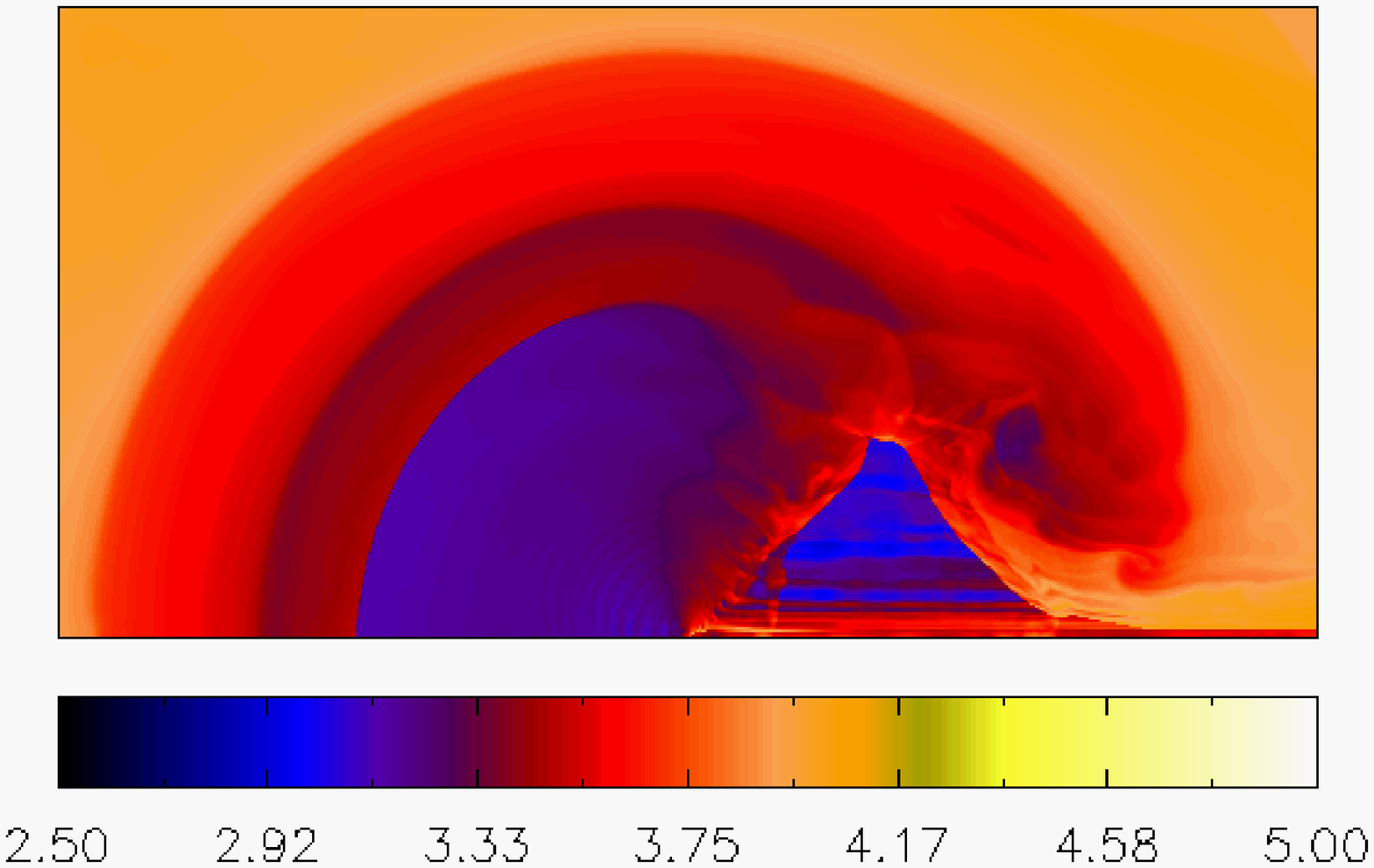}
\caption{ Photoevaporating minihalo snapshots of temperature for $10^7M_\odot$
minihalo encountered by a global I-front at $z=9$ from a source with 50,000 K 
black-body spectrum. Times shown are:
(left) 0.2 Myr, when I-front is still fast, R-type, with extended 
structure in IGM due to the finite photon mean free path clearly visible, and 
(right) 150 Myr, close to the complete evaporation, after which only 
a dark halo devoid of gas remains. Note complex flow structure which 
involves multiple shocks. Shaded isocontours indicate the values of $\log_{10}(T)$, as 
labeled on the color bar. 
\label{photoevap}
\vspace{-0.5cm}}
\end{figure}

I-fronts propagating outward from the ionizing sources encounter these
minihalos and get trapped inside them, eventually completely photoevaporating
all the gas. We have studied this process in detail, by means of
high-resolution adaptive-mesh-refinement (AMR) hydrodynamic and radiative 
transfer simulations 
\citep{2004MNRAS.348..753S,ISR05}. In Figure~\ref{photoevap} we show two 
snapshots of the photoevaporative flow's temperature structure at times 
$t=0.2$ Myr (during the fast, R-type phase), showing the relatively extended
I-front structure compared to the minihalo size, and $t=150$ Myr, close to the 
moment of complete evaporation of the minihalo, which shows the complicated 
structure of the photoevaporative flow, with multiple shocks (see papers for 
details). 
We have performed a large number of these high-resolution simulations (with finest 
grid up to $(r,z)=1024\times2048$ in 2-D axisymmetry) and found 
simple analytical fits to the total consumption of ionizing photons and 
evaporation times as functions of the minihalo initial mass, redshift, and 
the flux and spectrum of the ionizing source \citep{ISR05}. We showed
that minihalos can increase the global ionizing photon consumption during 
reionization by a factor of up to 2 compared to the mean IGM alone. 

\begin{figure}
 \includegraphics[width=2.6in]{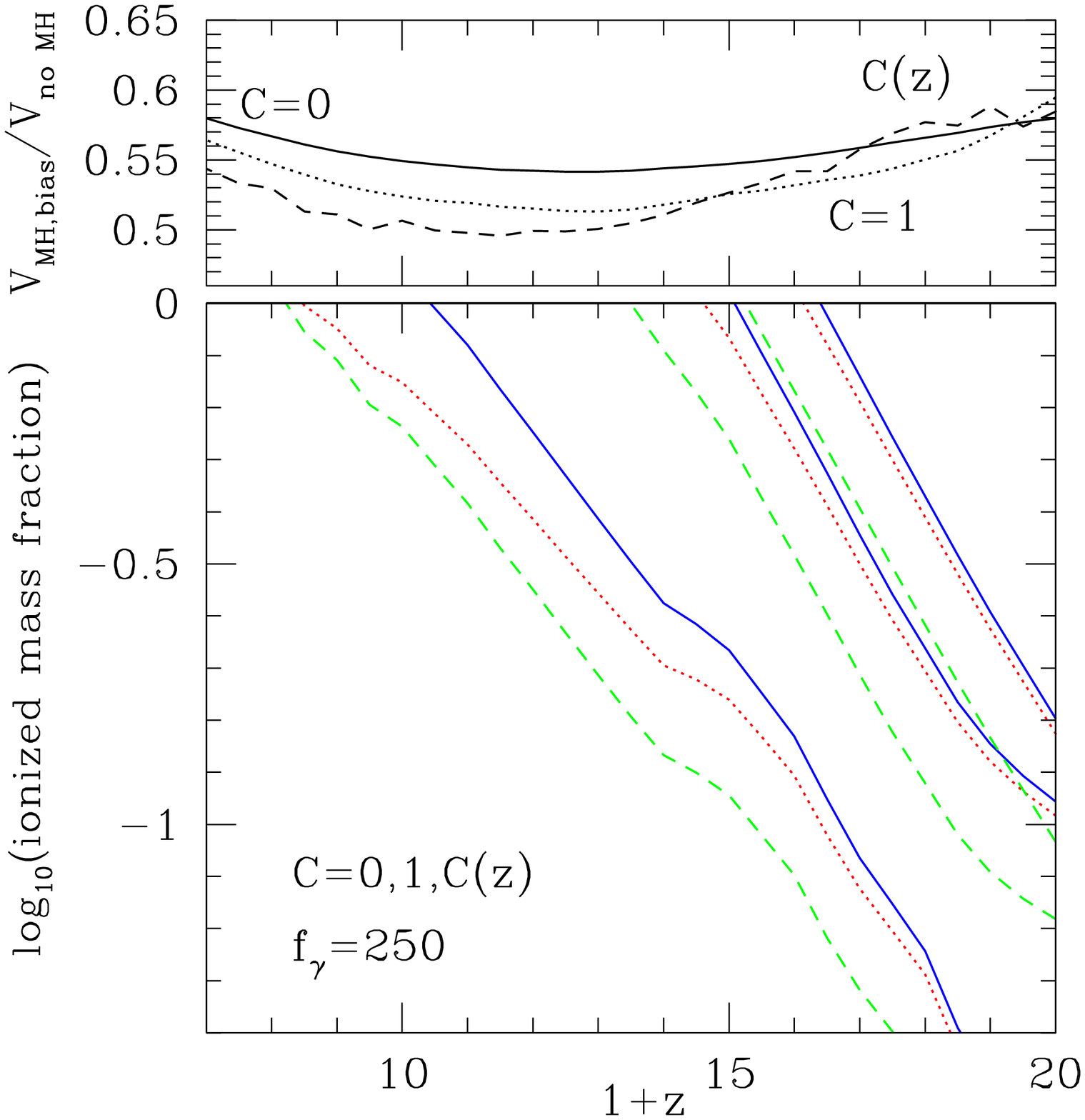}
 \includegraphics[width=2.7in]{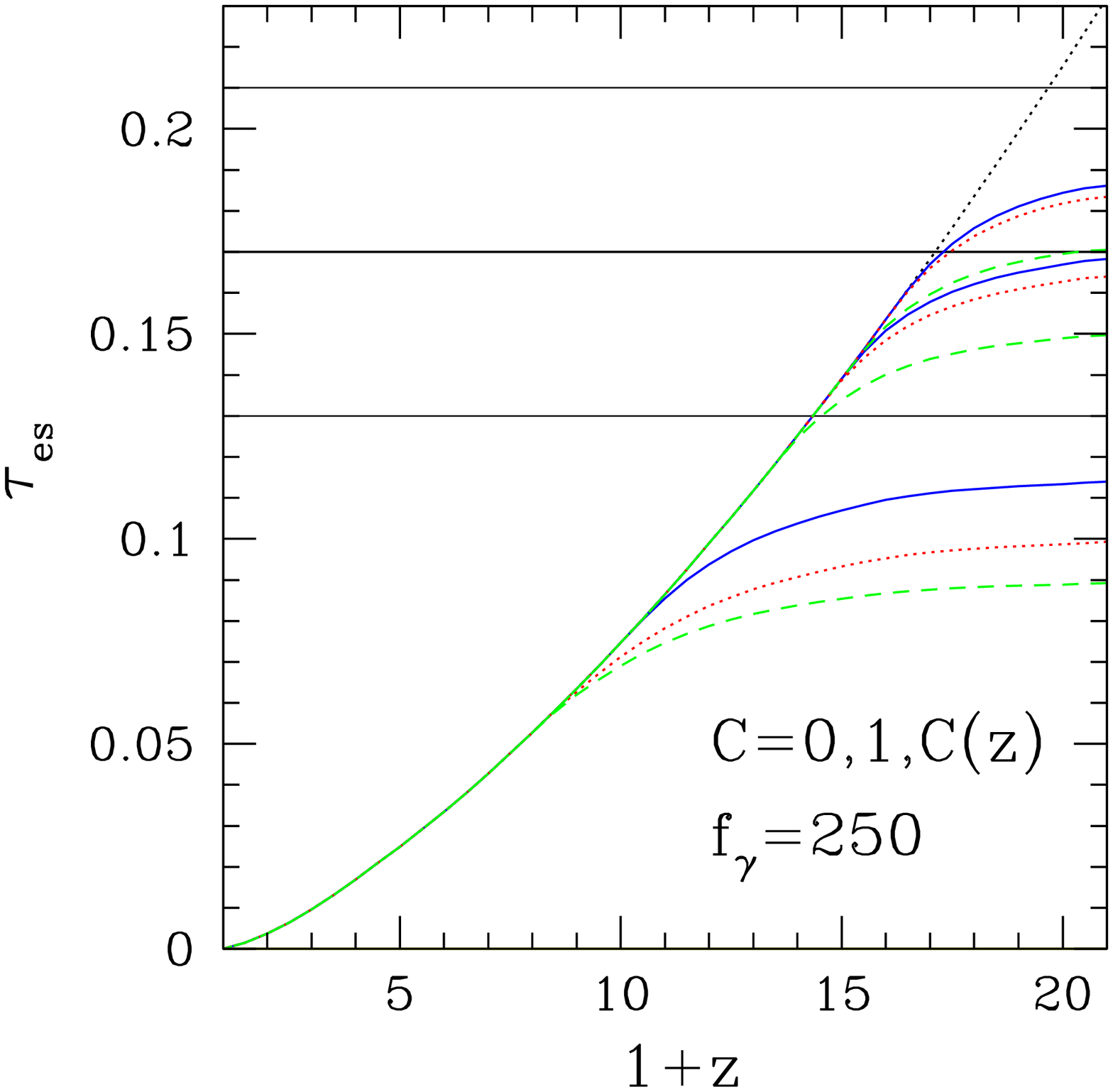}
  \caption{Global reionization. Sources with $f_\gamma=250$ and lifetime
  of $t_s=3$ Myr are assumed. (bottom panel) Decimal
  logarithm of ionized mass (or Lagrangian volume) fractions (i.e. 0
  corresponds to overlap) for the cases of no minihalos (solid),
  unbiased minihalos (dotted), and biased minihalos (dashed) for IGM
  clumping factors (top to bottom in each case) $C=0$, 1, and C(z)
  (clumped IGM). (upper panel) Ratios of the ionized volume fractions
  in the presence of {\em biased} minihalos and with no minihalos for $C=0, 1,C(z)$
\label{global_figs}
\vspace{-0.3cm}}
\end{figure}

\begin{figure}
 \includegraphics[width=2.7in]{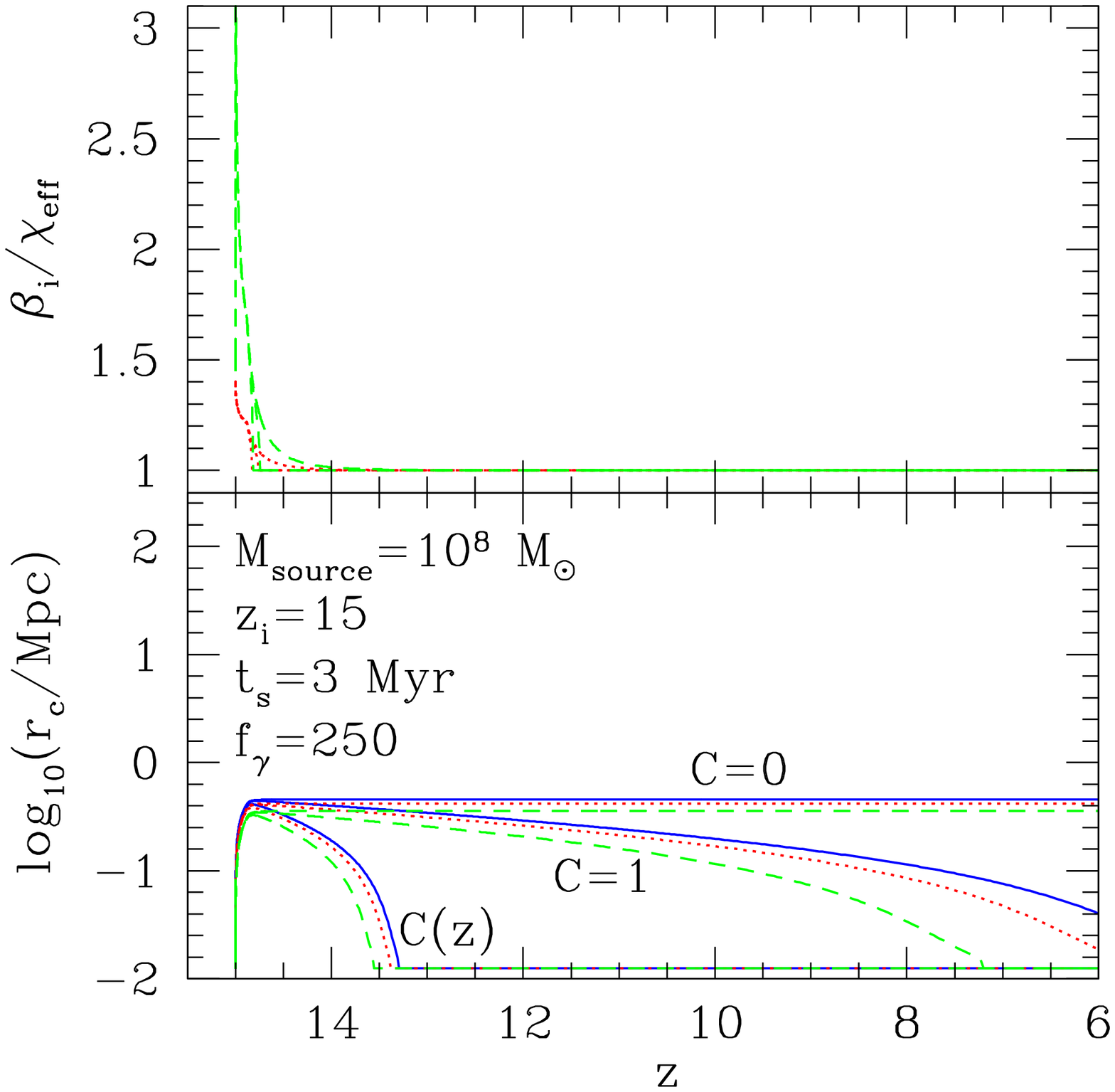}
 \includegraphics[width=2.7in]{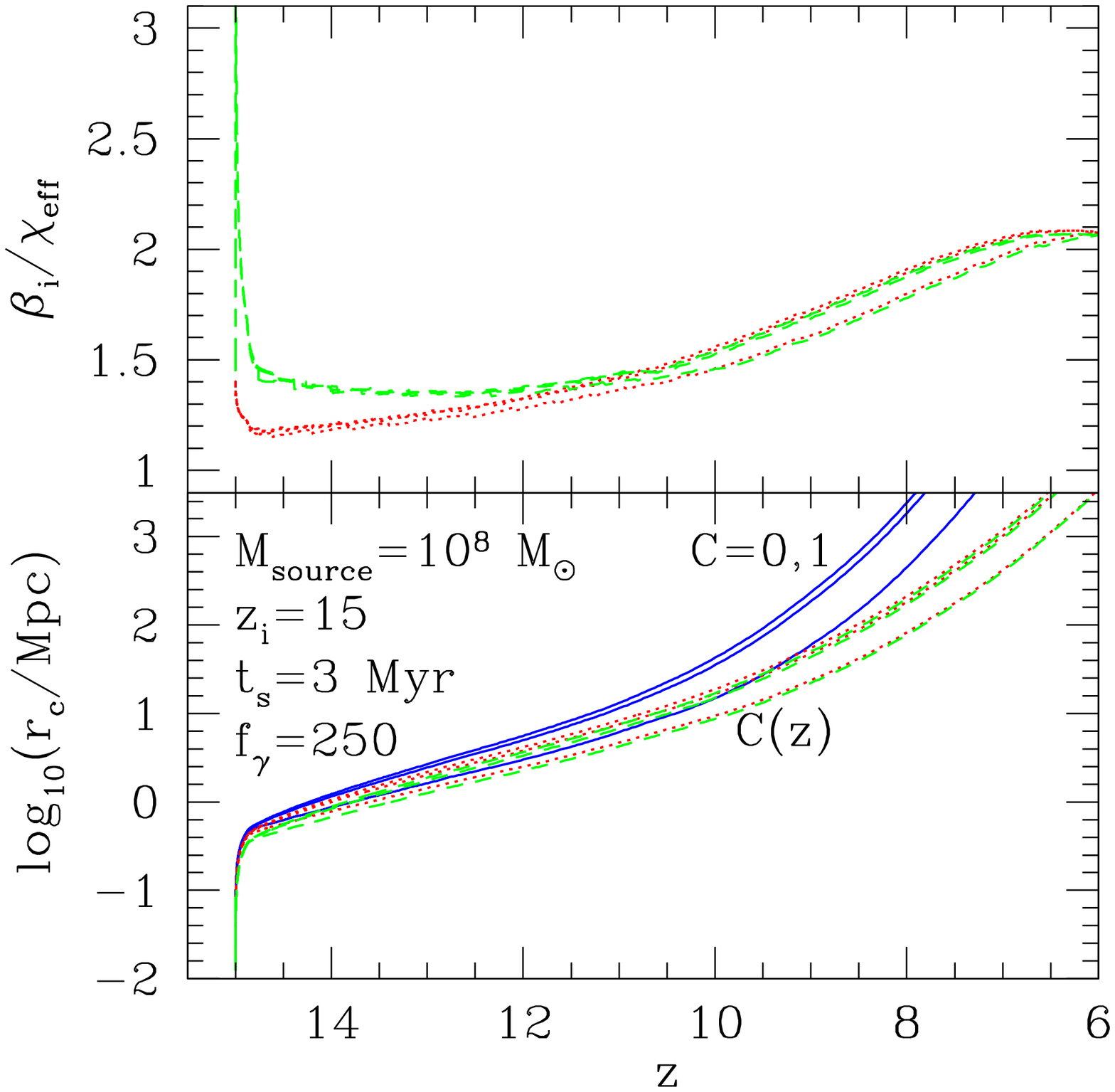}
  \caption{Evolution of an individual H~II region about a central
  source of mass $10^{8}M_\odot$ which turns on at redshift $z=15$, 
for (left) a single source, and (right) clustered multiple sources for $C=0,1, C(z)$.
{\em Top:} The correction factor 
$\beta_i/\chi_{\rm eff}$ due to minihalos for the number 
of ionized   photons consumed per atom that crosses the I-front, 
for biased (dashed) and  unbiased minihalos (dotted). 
{\em Bottom:} Comoving radius of the H~II region for no minihalos (solid), unbiased
  (dotted) and biased minihalos (dashed).
\label{source_clustering_figs}
\vspace{-0.5cm} }
\end{figure}

\begin{figure}
 \includegraphics[width=2.7in]{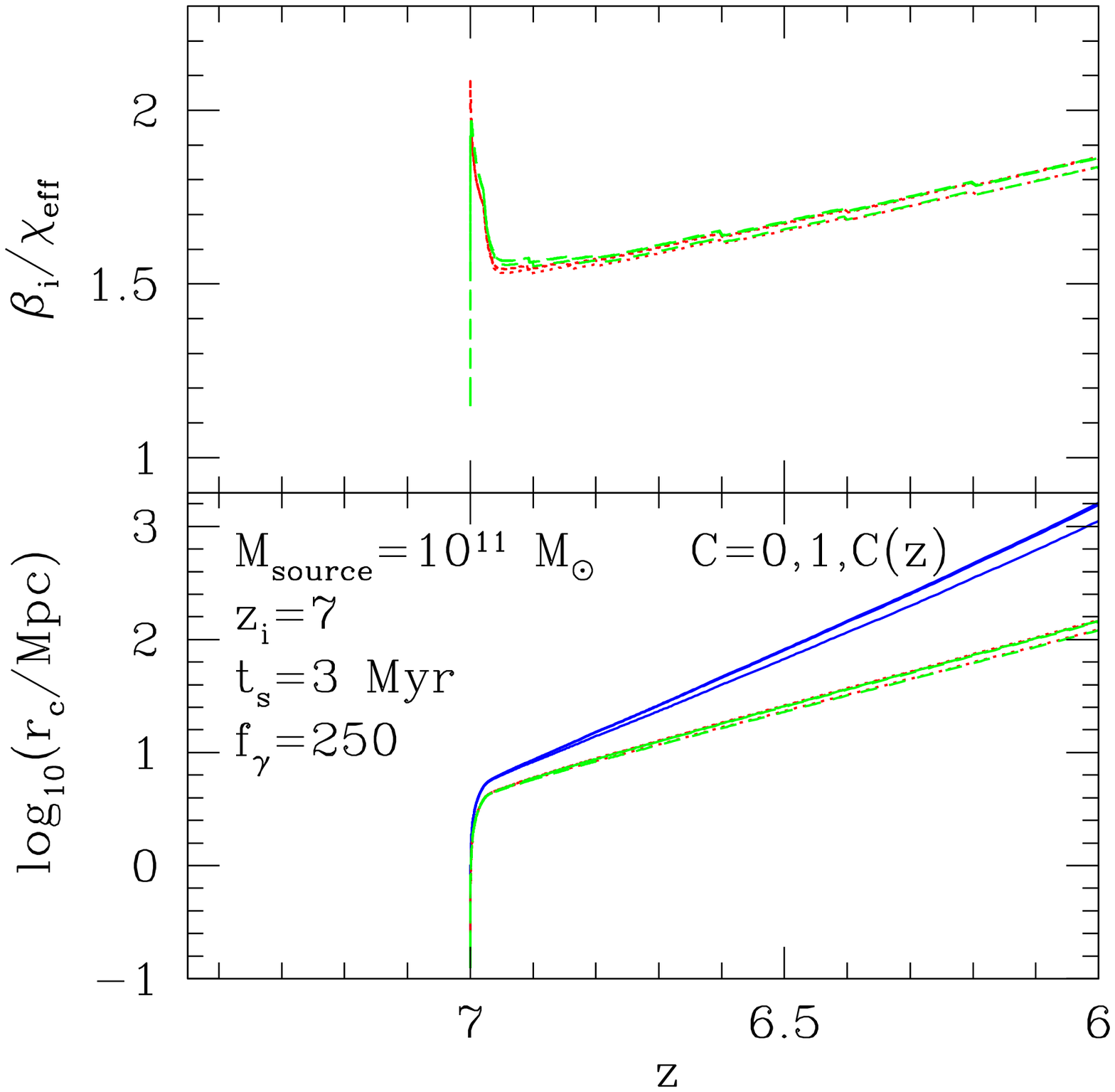}
 \includegraphics[width=2.7in]{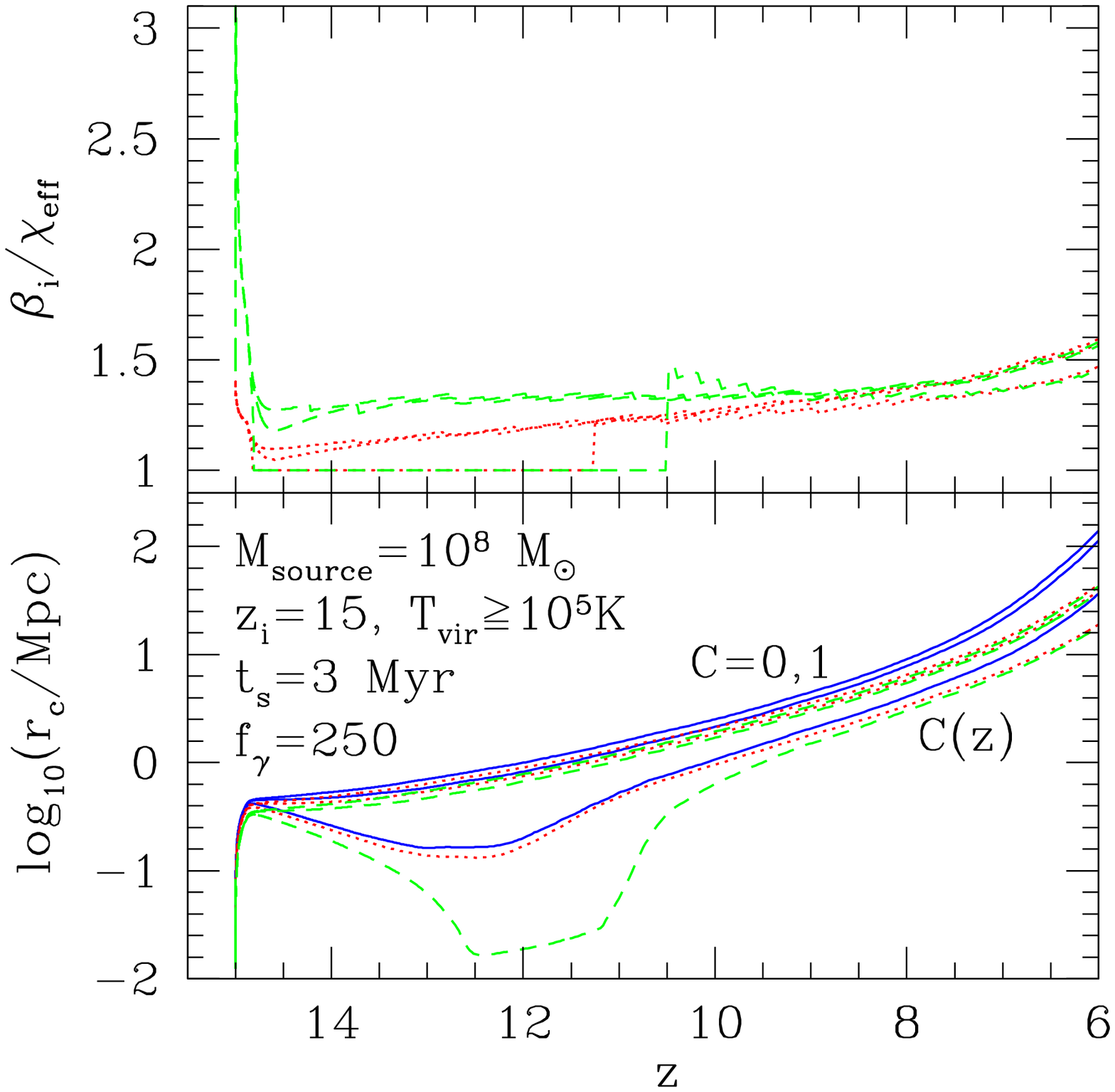}
  \caption{ Same as Fig.~\ref{source_clustering_figs}, for 
clustered sources, but (left)  a central source of mass $10^{11}M_\odot$ turning 
on at $z=7$, and (right) assuming that the formation of ionizing sources 
with virial temperatures below $10^5$K is suppressed within the H~II region.
\label{source suppr_figs}
\vspace{-0.5cm} }
\end{figure}

Using our fitting 
formulae to our detailed simulation results, we were able to modify the I-front 
propagation equations of \citet{1987ApJ...321L.107S} to properly reflect the 
increased consumption of ionizing 
photons due to the minihalos and the corresponding slowing down of the global
I-fronts \citep{ISS05}. We included the effects of infall and minihalo 
bias around the sources, the evolving mean IGM clumping and the dependence 
on the photon production efficiencies, 
spectra and lifetimes of the ionizing sources. We found that small-scale structures 
have a significant effect on the progress of reionization, slowing it down and 
extending it in time, which can help us understand the recent observations by the 
Wilkinson Microwave Anisotropy Probe satellite, which point to an early and extended 
reionization epoch (Fig.~\ref{global_figs}). 
More recently (work in progress) we extended our formalism to account for the 
clustering of the ionizing sources. We find that the clustering of source
halos in time and space leads to multiple sources in each H~II region, which 
dramatically 
changes the numbers and sizes of the ionized regions and hence the overall topology 
of reionization (Fig.~\ref{source_clustering_figs} and Fig.~\ref{source suppr_figs}, 
left). H~II regions around short-lived individual sources are small (less than 
1 Mpc comoving 
around the smaller sources) and quickly recombine and shrink in size once the source 
ends its life, forming so-called ``relic H~II regions'', but when source halo 
clustering is taken into account the H~II regions expand continuously, reaching 
sizes of up to tens and hundreds of comoving Mpc. In that case, relic H~II regions 
do not generally form. We also study the consequences of the suppression of the 
formation of smaller sources inside the already-ionized 
regions (due to increased Jeans mass there, in correspondence to the higher gas 
temperature of $\sim \rm few\times10^4$ K) (Fig.~\ref{source suppr_figs}, right). We 
find that sufficiently strong suppression can have quite dramatic effects, 
decreasing (in the particular high-redshift source case we consider here) the 
size of the H~II region by an order of magnitude, and leading to a temporary 
formation of a small relic H~II region. However, for lower-redshift sources this
suppression has much smaller consequences since the larger ionizing sources (whose 
formation is not affected by this suppression), which were very rare at high-z, become 
more abundant and start playing a more important role in the process of reionization.  

\section{A new photon-conserving method for transfer of ionizing radiation}

\begin{figure}
 \includegraphics[width=2.7in]{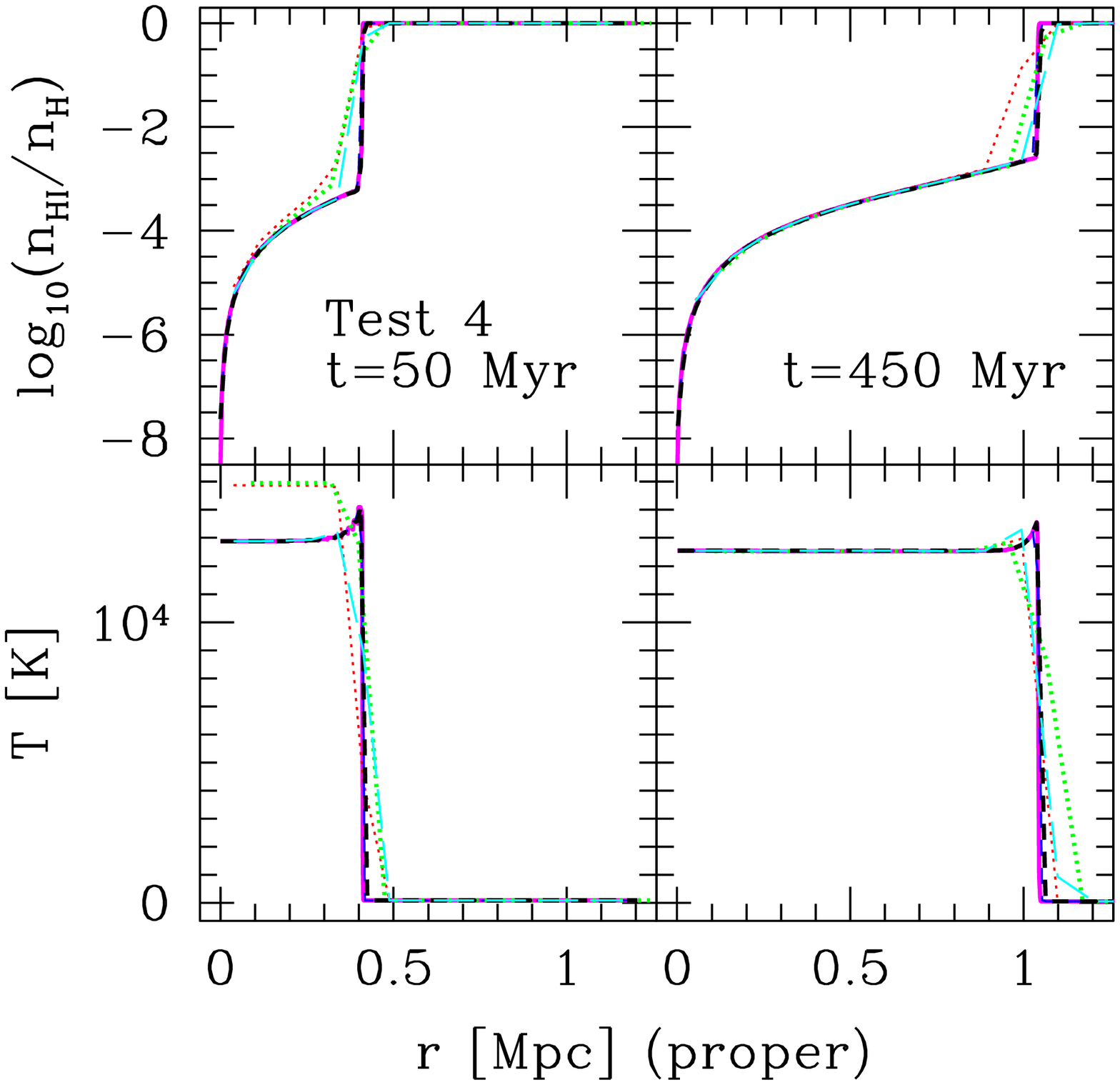}
 \includegraphics[width=2.4in]{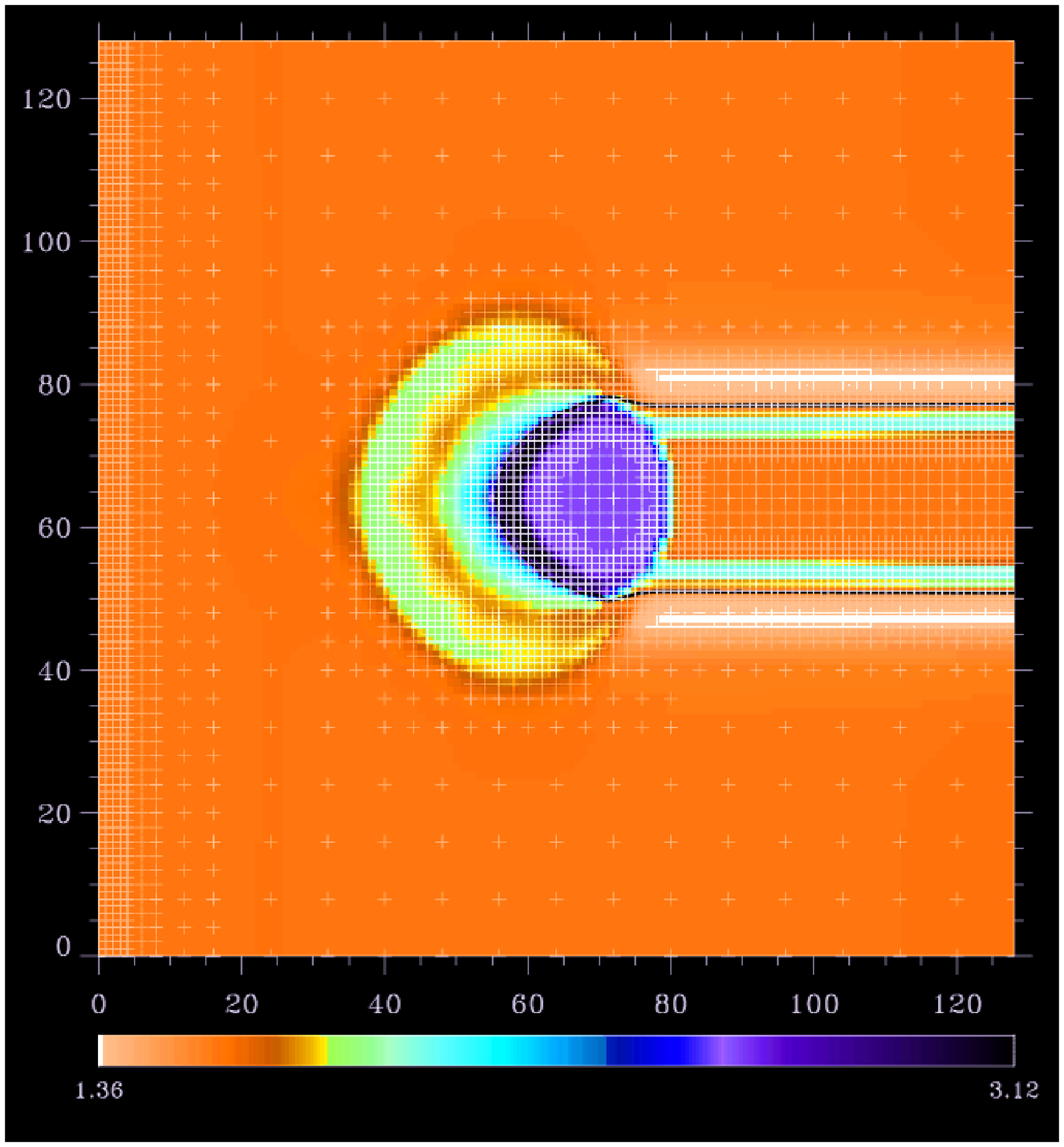}
  \caption{Photon-conserving RT. (left) Neutral hydrogen (top) and temperature 
(bottom) profiles for cosmological I-front starting at $z=9$. The left panels show 
time $t=50$~Myrs after turn-on ($=1$ time step in coarse temporal resolution case, 
and 10 time steps in the high temporal resolution case), while right panels show 
  solution at $t=450$~Myrs (=$3.5t_{\rm rec}$). Thin lines correspond to 1-D results 
with 16 cells, 10 time-steps (dotted, red), 16 cells, 100 time-steps (long-dashed, 
cyan), and 128 cells, 100 time-steps (short-dashed, blue). Thick lines correspond 
to 3-D results with $32^3$ cells, 10 time-steps (dotted, green) and $256^3$ cells, 
100 time-steps (short-dashed, black). Finally, the reference ``exact'' solution 
(1-D, 1024 cells, 100 time-steps) is shown by thick, solid, magenta line. (right)
 3-D, AMR gasdynamics and radiative transfer simulation of photoevaporation of a
dense gas clump calculated with a 5-level adaptive mesh. Snapshot shows
$\log_{10}(n)$, gas number density in central $xy$ plane. White crosses 
indicate computational mesh, while black shows I-front position.
\label{RT_code_figs}
}
\end{figure}

In the last few years
there has been an intense development of new numerical radiative transfer
methods for cosmology \citep[e.g.][]{1999ApJ...523...66A, 2001NewA....6..437G,
 2002ApJ...572..695R,2003MNRAS.345..379M}. 
However, most of the existing methods are not efficient enough to be directly 
coupled to gas- and N-body 
dynamics, which limits the scope of their applications. One common problem is that the
fast, R-type I-front propagation timescale (sometimes approaching the light-crossing 
time) is much shorter than the fluid dynamical and gravitational times, leading to the 
necessity of very small time-steps to track the I-fronts  correctly. In addition, 
the bound-free opacity of the neutral gas is very high, so algorithms for spatial 
differencing the transfer equation often require very small cell sizes, and this 
implies very small time steps even when the I-fronts are subsonic. Our photoevaporation 
simulations described above \citep{2004MNRAS.348..753S,ISR05} utilized an AMR Eulerian 
hydro code in 2-D axisymmetry, capable of handling a large range of scales with high 
resolution efficiently enough that we were able to take very small time-steps and very 
small cell sizes where needed. However, to generalize these ``zoomed-in'' calculations 
to 3-D and a simulation volume large enough to encompass both small- and large-scale
structure, it is necessary to develop a more sophisticated radiative transfer algorithm.
We have recently developed such a method, which accurately
follows both fast and slow I-fronts without need to adopt such small time-steps or 
cell sizes. We achieved this by matching the ionization rates and flux attenuation
along rays to ensure photon conservation, and time-averaging the neutral column 
densities 
per cell per step to give the correct I-front position and speed even for large 
time-steps and cell-widths \citep{methodpaper,hydropaper} (see papers for 
details). These characteristics make our method very efficient for direct coupling with
gasdynamic and N-body codes. We have tested the method in detail, on both fixed and 
adaptive grids, as a stand-alone radiative transfer code as well as coupled to
AMR gasdynamics (for some sample results see 
Fig.~\ref{RT_code_figs}).



\end{document}